%% file: main.tex
\documentclass[11pt]{article}
\usepackage[T1]{fontenc}
\usepackage[utf8]{inputenc}
\usepackage{lmodern}
\usepackage[margin=1in]{geometry}
\usepackage{amsmath,amssymb,amsthm,mathtools}
\usepackage{booktabs,array,longtable}
\usepackage{microtype}
\usepackage{xurl}
\usepackage[hidelinks]{hyperref}
\newcommand{\paperTitle}{A Numerical Refinement of an Exact Discrete Fourier Transform Bound}
\newcommand{\paperAuthor}{Zhihao Yao and Sichao Liu}
\newcommand{\artifactDOI}{\texttt{\detokenize{https://doi.org/10.5281/zenodo.23843356}}}
\hypersetup{pdftitle={\paperTitle},pdfauthor={\paperAuthor}}
\newtheorem{theorem}{Theorem}[section]
\newtheorem{lemma}[theorem]{Lemma}
\newtheorem{proposition}[theorem]{Proposition}
\newtheorem{corollary}[theorem]{Corollary}

\newcommand{\paid}{\mathcal P}
\newcommand{\target}{\mathcal T}
\newcommand{\mom}[2]{M_{#1}(#2)}

\DeclareMathOperator{\DFT}{DFT}
\title{A Numerical Refinement of an Exact\\Discrete Fourier Transform Bound}
\author{Zhihao Yao\\[3pt]\small New Jersey Institute of Technology\\Newark, USA
  \and Sichao Liu\\[3pt]\small Independent Researcher\\West Orange, USA}
\date{}

\begin{document}
\maketitle

\begin{abstract}
We refine the numerical bounds in Kartal's instantiation of Sussman's exact
discrete Fourier transform construction. For every integer $n\geq1$, the
resulting algorithm computes the length-$n$ transform with work
$O(n(\log n)^{1-\delta})$, where $\delta=0.0030201659$.
The bound holds in an exact complex-arithmetic RAM model with a supplied
canonical root of unity. This final exponent saving exceeds the earlier
certified parameter $C_0=0.00302016$ by $5.9\times10^{-9}$.
We retain the underlying circuit, tighten rational bounds for sums of
powers of its block and target ranks, and increase the fixed tensor power
from $7760$ to $11640$. These bounds establish the inequalities for
Sussman's recursion at the intermediate certificate parameter
$C=0.00302016595$. The gap $C-\delta$ absorbs the overhead of recursion
and Fourier reduction. The resulting transform theorem is proved in Lean
and checked against the reference problem statement.
\end{abstract}

\section{Introduction}

We consider the arithmetic work needed to compute the discrete Fourier
transform for every integer input length $n\geq1$. Cooley and Tukey's factorization
gives the familiar
$O(n\log n)$ bound, while Bluestein's reduction extends this order of growth
to arbitrary lengths~\cite{cooley1965,bluestein1970}. More recent work has
reduced leading constants: Alman and Rao obtain
$\frac{15}{4}n\log_2 n+O(n)$ real arithmetic operations for lengths that are
powers of two~\cite{alman2023}.

Another line of work studies a saving in the exponent of the logarithm under
exact complex arithmetic. OpenAI's construction for arbitrary lengths gives such a
saving in a uniform RAM model with charged scalar preparation and a supplied
root of unity~\cite{openai2026}. Subsequent finite circuits and formalizations
have increased the explicit saving. The present work concerns the
instantiation by Kartal~\cite{kartal2026}, which uses Sussman's mixed-target
recursion~\cite{sussman2026} and the controller and tensor constructions
developed by Davis and the integer-multiplication circuit
community~\cite{davis4416,davis4440,davis4470}.

The exponent in this construction is controlled by a finite set of
inequalities involving the sizes and counts of its component computations.
We call the exact numerical bounds that verify these inequalities a
\emph{numerical certificate}. Kartal's submission states a final Fourier
saving of $0.0030103$, while its certificate establishes the stronger
parameter $C_0=4719/1562500=0.00302016$~\cite{kartal2026}.
The recursion and Fourier reduction permit any fixed saving
$0<\delta<C_0$: the gap below $C_0$ absorbs their asymptotic overhead.
Thus, for example, a saving of $0.00302$ already follows from the earlier
argument. We improve the numerical certificate to obtain a saving beyond
$C_0$.

\begin{samepage}
\begin{theorem}\label{thm:main}
In the exact complex-arithmetic RAM model described in
Section~\ref{sec:model}, there are fixed deterministic programs that compute
the length-$n$ discrete Fourier transform for every integer $n\geq1$, with work
\[
  W(n)=O\!\left(n(\log n)^{1-\delta}\right),
  \qquad \delta=\frac{30201659}{10^{10}}=0.0030201659.
\]
The work bound includes the root-order computation, one supplied-root
operation, scalar preparation, and array and index operations. Integer values
and addresses are bounded by a fixed polynomial in $n$.
\end{theorem}
\end{samepage}

The improvement is a numerical refinement of the existing algorithm. We
use the same raw circuit, the same sizes and counts of its component
computations, and the same reduction to the Fourier transform. The changes are tighter rational
estimates of the rank moments and a larger fixed tensor power. 
Our final DFT saving is $\delta=0.0030201659$, exceeding the earlier
certified parameter $C_0=0.00302016$ by $5.9\times10^{-9}$.
Earlier circuit analysis placed the raw moment balance point near
$0.003020165977$.
Our final saving remains below this value to accommodate the overhead
of the tensor construction, recursion, and Fourier reduction.
Conditional integer-multiplication constructions
had also used rates close to that point~\cite{hexapode4568,chris4707}.

The proof below gives the finite arithmetic behind the refinement and then
applies Sussman's recursion and the Fourier reduction. The accompanying
Lean proof verifies the resulting transform algorithm and its work bound
in the reference model.

\section{Computational model and the finite unit}\label{sec:model}

For $n\geq1$, write $\zeta_n=\exp(2\pi i/n)$. We use the positive-sign,
unnormalized transform
\begin{equation}\label{eq:dft}
  (\DFT_n x)_j=\sum_{k=0}^{n-1}\zeta_n^{jk}x_k,
  \qquad 0\leq j<n.
\end{equation}
All logarithms in complexity bounds are natural logarithms.

We retain OpenAI's exact RAM model~\cite{openai2026}. A fixed program first
computes a positive integer $D(n)$ from $n$. A root primitive then supplies
the exact complex number $\zeta_{D(n)}=\exp(2\pi i/D(n))$, after which a
second fixed program computes~\eqref{eq:dft}. The order $D(n)$ may differ
from the transform length $n$. The formal specification requires
$D(n)<1024n^3$ and bounds all integer values and addresses by $(n+2)^c$ for
one constant $c$. Its cost includes both programs and the root primitive.
Exact scalar arithmetic and integer-word, array, and index operations are
charged according to the RAM semantics. Scalar coefficients have unrestricted
magnitude. The work bound counts exact arithmetic and RAM operations.

The finite unit used here is Kartal's P16A8Z instance~\cite{kartal2026}.
It combines a 16-pair controller, with prefix parameter eight, and an
18-stage geometry with 145 label coordinates. Its supplier comes from the
patched circuit in Davis's joint-helper construction~\cite{davis4470}.
In the cost analysis, a \emph{charged block} is a component computation
whose work is counted, and a \emph{target job} specifies a computation
that the circuit carries out. The \emph{rank} of a block or job is the
number of coordinate axes on which it acts. After dividing out a common
multiplicity factor, the ranks and their counts form the two multisets
in Table~\ref{tab:inventory}. For example, the entry $(2,6984)$ describes
$6984$ charged blocks of rank two. We use the same two multisets as
Kartal's construction.

\begin{table}[tb]
\centering
\caption{Ranks and counts of charged blocks and target jobs in the raw
P16A8Z construction, with the common multiplicity factor omitted.}
\label{tab:inventory}
\input{data/inventories}
\end{table}

For a finite multiset $A$ of positive integers, define its moment by
\[
  \mom{t}{A}=\sum_{r\in A}r^t,
\]
where the sum counts multiplicity. Let $\paid$ and $\target$ be the charged-block and
target multisets, and abbreviate
$P_t=\mom{t}{\paid}$ and $T_t=\mom{t}{\target}$. Direct summation gives
\begin{equation}\label{eq:mass}
  P_1=239134,\qquad T_1=239886,\qquad |\target|=10928.
\end{equation}
The recursion compares $P_t$ with $T_t$ at an exponent $t<1$. It also needs
every charged-block rank to be smaller than every target rank. The raw multisets
do not have this separation: some block ranks exceed the smallest target
rank. Tensor products, target completion, and block splitting establish
it in Section~\ref{sec:tensor}.

\section{The refined numerical bounds}\label{sec:certificate}

Fix the exact rational numbers
\begin{equation}\label{eq:parameters}
  C=\frac{60403319}{20000000000}=0.00302016595,
  \qquad t=1-C,\qquad
  \varepsilon=\frac{27}{10^{12}}.
\end{equation}
Our first task is to prove $P_t\leq(1-\varepsilon)T_t$. The gap is small
enough that the earlier exponential estimates lose the required precision.
The following elementary bounds suffice.

\begin{lemma}\label{lem:exp}
Let $S_6(x)=\sum_{j=0}^{5}x^j/j!$. For $x\geq0$,
$S_6(x)\leq e^x$. For $0\leq x\leq1$,
\begin{equation}\label{eq:exp}
  e^x\leq S_6(x)+\frac{7}{4320}x^6.
\end{equation}
\end{lemma}
\begin{proof}
The lower bound discards nonnegative terms of the exponential series.
For the upper bound, its tail starts with $x^6/6!$. Each subsequent term
is at most $1/7$ times the preceding term when $x\leq1$. The tail is
therefore at most $(x^6/720)/(1-1/7)=7x^6/4320$.
\end{proof}

We normalize at the largest target rank, 145. For a rank $r$ with
multiplicity $m$, its contribution becomes
\begin{equation}\label{eq:normalized}
  m\left(\frac r{145}\right)^{1-C}
  =\frac{mr}{145}\exp\!\left(C\log\frac{145}{r}\right).
\end{equation}
Every argument of the exponential lies in $[0,0.016]$. Rational bounds for
$\log(145/r)$ can therefore be substituted into Lemma~\ref{lem:exp} with
little loss. Appendix~\ref{app:bounds} gives the bounds for every term,
together with the finite-series tests that certify the logarithms.

\begin{samepage}
\begin{proposition}\label{prop:moment}
At the parameters in~\eqref{eq:parameters},
\begin{align}
  \frac{P_t}{145^t}&\leq
       \frac{1661483013014771}{10^{12}}=:U,
       \label{eq:upper}\\
  \frac{T_t}{145^t}&\geq
       \frac{1661483013061537}{10^{12}}=:L.
       \label{eq:lower}
\end{align}
Consequently, $P_t/T_t\leq1-\varepsilon$.
\end{proposition}
\end{samepage}
\begin{proof}
For each charged-block rank, use an upper logarithm bound in~\eqref{eq:normalized}
and then~\eqref{eq:exp}. For each target rank, use a lower logarithm bound
and the lower polynomial $S_6$. The rank-145 target term is exactly 224.
Adding the respective rational enclosures in
Table~\ref{tab:bounds} gives~\eqref{eq:upper} and~\eqref{eq:lower}.
Both sums are positive, and exact rational arithmetic gives
\[
  1-\frac UL=\frac{46766}{1661483013061537}
  >\frac{27}{10^{12}}.
\]
Since $P_t/T_t\leq U/L$, the stated gap follows.
\end{proof}

The completion and splitting estimates also use the scaled target moment
$20^C T_t$. Applying the same upper exponential bound gives
\begin{equation}\label{eq:scale}
  \left(\frac{145}{20}\right)^C
  \leq E:=\frac{251500221699}{250000000000}.
\end{equation}
Hence
\begin{equation}\label{eq:scaled-target}
  20^C T_t
  =\frac{145}{(145/20)^C}\frac{T_t}{145^t}
  \geq\frac{145L}{E}.
\end{equation}
Set
\begin{equation}\label{eq:ratios}
  \gamma=\frac{99857}{100000},\qquad
  \eta=\frac{9583}{10000}.
\end{equation}
The rational inequalities
\[
  239134\leq\gamma\frac{145L}{E},\qquad
  10928\cdot21\leq\eta\frac{145L}{E}
\]
and~\eqref{eq:mass} imply
\begin{equation}\label{eq:tail-ratios}
  \frac{P_1}{20^C T_t}\leq\gamma,
  \qquad
  \frac{21|\target|}{20^C T_t}\leq\eta.
\end{equation}
All numbers used in these comparisons are rational.

\section{A separated tensor construction}\label{sec:tensor}

We use the completion and splitting argument already present in the
earlier construction~\cite{davis4416,kartal2026}. The moment estimate is
recalled here to make the role of the larger tensor power explicit.

For multisets $A,B$, let $A\otimes B$ contain every product $ab$, with
multiplicity, for $a\in A$ and $b\in B$. Moments multiply:
\begin{equation}\label{eq:product}
  \mom{t}{A^{\otimes v}}=\mom{t}{A}^{v}.
\end{equation}
In the $v$-fold tensor unit, choose positive integers $J<B$ as the maximum
block rank and minimum target rank. A target of rank $K<B$ is completed
to rank $B$ by adding a charged block of rank $B-K$. Targets of rank at
least $B$ are unchanged. Each charged block of rank $R>J$ is then split into blocks of rank
$J$ and one remainder block, if needed. Write $\paid'$ and $\target'$ for
the resulting multisets. Every charged-block rank is at most $J$, and every
target rank is at least $B$.

\begin{lemma}[Completion and splitting]\label{lem:split}
Let $0\leq t\leq1$ and $v\geq1$. Then
\begin{equation}\label{eq:split}
 \mom{t}{\paid'}\leq
 P_t^v+2J^{t-1}P_1^v+2|\target|^v B J^{t-1},
 \qquad
 \mom{t}{\target'}\geq T_t^v.
\end{equation}
\end{lemma}
\begin{proof}
When $R>J$, the sum of the $t$th powers of its pieces is at most
\[
  \left\lfloor\frac RJ\right\rfloor J^t+(R\bmod J)^t
  \leq RJ^{t-1}+J^t\leq2RJ^{t-1}.
\]
For $R\leq J$, the contribution is $R^t$. Thus a bound valid in both
cases is $R^t+2RJ^{t-1}$. Summing over the charged blocks in the tensor unit and
using~\eqref{eq:product} gives the first two terms in~\eqref{eq:split}.
There is at most one completion block for each target in
$\target^{\otimes v}$, and its rank is less than $B$. After splitting,
its contribution is at most $2BJ^{t-1}$. If its rank is already at most
$J$, its contribution is at most $J^t\leq BJ^{t-1}$. There are $|\target|^v$ possible
completion blocks. Finally, completion only increases target ranks, so
their $t$-moment is at least $T_t^v$.
\end{proof}

Take $J=20^v$ and $B=21^v$. At $t=1-C$, division by $T_t^v$ in
Lemma~\ref{lem:split} reduces the desired comparison to
\begin{equation}\label{eq:three-terms}
  \left(\frac{P_t}{T_t}\right)^v
  +2\left(\frac{P_1}{20^C T_t}\right)^v
  +2\left(\frac{21|\target|}{20^C T_t}\right)^v<1.
\end{equation}
By Proposition~\ref{prop:moment} and~\eqref{eq:tail-ratios}, it is enough
to prove $(1-\varepsilon)^v+2\gamma^v+2\eta^v<1$.

\begin{proposition}\label{prop:tensor}
For $v=11640$, the completed and split tensor unit has
$\mom{t}{\paid'}<\mom{t}{\target'}$, and every charged-block rank is smaller
than every target rank.
\end{proposition}
\begin{proof}
Exact rational exponentiation gives
\[
  \gamma^{97}\leq\frac{544}{625},\qquad
  \left(\frac{544}{625}\right)^5\leq\frac12,
  \qquad \eta^{40}\leq\frac12.
\]
Since $11640=485\cdot24=40\cdot291$,
$\gamma^{11640}\leq2^{-24}$ and
$\eta^{11640}\leq2^{-291}$. Also, Bernoulli's inequality gives
$(1-\varepsilon)^v\leq(1+v\varepsilon)^{-1}$. Therefore
\begin{equation}\label{eq:final-rational}
 (1-\varepsilon)^{11640}+2\gamma^{11640}+2\eta^{11640}
 \leq\frac{1}{1+11640\varepsilon}+2^{-23}+2^{-290}<1.
\end{equation}
The last strict inequality is rational; its slack exceeds
$195/10^9$. Equations~\eqref{eq:split}--\eqref{eq:three-terms} now give
the moment comparison. Rank separation follows from
$20^{11640}<21^{11640}$.
\end{proof}

The earlier certificate used $v=7760$. At the new rate the available
moment gap is much smaller, so increasing $v$ lets the two tail terms
fall below the accumulated gain in the first term.

\begin{proof}[Proof of Theorem~\ref{thm:main}]
The P16A8Z certificate realizes the raw rank multisets by an exact
circuit~\cite{kartal2026}. Its tensor construction realizes the completed
and split multisets used above. Proposition~\ref{prop:tensor} supplies
the moment comparison and rank separation required by the mixed-target
recursion of Sussman~\cite{sussman2026}. That recursion gives the kernel
program used in the reduction to arbitrary transform lengths at any fixed exponent strictly
above $1-C$.

Choose the intermediate saving
\[
  \beta=\frac{3020165925}{10^{12}}=0.003020165925.
\]
Then $\delta<\beta<C$. With kernel exponent $1-\beta$, the
Fourier reduction, including its preparation costs, gives
\[
  W(n)=O\!\left(n(\log n)^{1-\beta}(\log\log n)^2\right).
\]
For the positive constant $\beta-\delta$,
$(\log\log n)^2=o((\log n)^{\beta-\delta})$.
This proves the asserted work bound. The fixed-program, root-order, and
polynomial integer bounds are supplied by the same reduction.
\end{proof}

\begin{corollary}
In the corresponding exact RAM model allowing multiplication of the two
input data streams, the linear convolution of two vectors of length $n$
can be computed in $O(n(\log n)^{0.9969798341})$ work.
\end{corollary}
\begin{proof}
Apply the convolution reduction to transform length $2n-1$.
The scalar and pointwise multiplication work is included in that
reduction's cost bound, and $2n-1=\Theta(n)$.
\end{proof}

\section{Verification and numerical comparison}\label{sec:verification}

Theorem~\ref{thm:main} and the convolution corollary are verified in
Lean 4.34.1. Using the development wrapper of Comparator~\cite{comparator},
a statement comparison checked the transform theorem against the reference exact-DFT
specification at the new rational exponent and rechecked the proof with
Lean's default kernel. The dependency audit reports propositional
extensionality, classical choice, and quotient soundness for the transform,
convolution, and finite tensor moment theorems. The artifact records these
checks and the source changes to the numerical bounds, tensor power, and
exponent parameters.

An independent calculation checks the numerical inequalities using exact
rational interval arithmetic. It uses an atanh series for logarithms and a separately
implemented exponential-series bound, rather than the generator's
logarithm certificates. This check covers the raw moment gap, both tail
ratios, and~\eqref{eq:final-rational}.

The interval calculation also resolves the raw balance point more closely.
Let $F(c)=P_{1-c}-T_{1-c}$. With
\[
 c_-=0.00302016597701562048,\qquad
 c_+=0.00302016597701562050,
\]
the certified signs are $F(c_-)<0<F(c_+)$. Continuity therefore places
a zero in $(c_-,c_+)$. The same calculation gives
$F(0.003020166)>0$, so that parameter fails the raw moment test.

The final Fourier saving remains strictly below this interval because
the proof uses slack at both the tensor and Fourier stages. The large
fixed tensor power also affects the hidden constants in the work bound.

\section*{Acknowledgments}
OpenAI ChatGPT Codex was used throughout the research and preparation of
this paper. Its contributions included analysis of prior constructions,
exploration of candidate improvements, derivation of the numerical bounds,
development of Lean proofs and verification scripts, execution and analysis
of verification checks, preparation of the reproducibility artifact, and
drafting and revision of the manuscript.
This work was supported in part by a grant of access to OpenAI models through the ChatGPT for Academic Researchers program.

\section*{Artifact availability}
The accompanying artifact contains the Lean development, exact rational
certificates, an independent arithmetic checker, source provenance, and
verification logs. Zenodo DOI: \artifactDOI.

\bibliographystyle{plain}
\bibliography{references}

\appendix
\section{Finite rational bounds}\label{app:bounds}

Table~\ref{tab:bounds} lists the logarithm and moment enclosures used in
Proposition~\ref{prop:moment}. Every displayed decimal in the table is an
exact rational with denominator dividing $10^{12}$. The multiplicity $m_r$
for each row is given in Table~\ref{tab:inventory}.

For $N\geq1$, put
\[
 S_N(x)=\sum_{j=0}^{N-1}\frac{x^j}{j!},\qquad
 U_N(x)=S_N(x)+\frac{N+1}{N\,N!}x^N.
\]
For $0\leq x\leq1$, the same tail argument as in
Lemma~\ref{lem:exp} gives $S_N(x)\leq e^x\leq U_N(x)$.
An upper logarithm entry $\ell_r$ is certified by
$145/r\leq S_N(\ell_r)$; the lower-series inequality holds for all
$\ell_r\geq0$. A lower logarithm entry is certified by
$U_N(\ell_r/8)^8\leq145/r$. In the latter case,
$0\leq\ell_r/8\leq1$, and the test implies $e^{\ell_r}\leq145/r$.
The charged-block moment bound $b_r$ satisfies
$\frac{m_rr}{145}U_6(C\ell_r)\leq b_r$.
The target bound satisfies
$b_r\leq\frac{m_rr}{145}S_6(C\ell_r)$.
All these tests reduce to integer arithmetic after clearing denominators.

For~\eqref{eq:scale}, use the upper logarithm bound
$\log(145/20)\leq1.981001468867$, certified by
$145/20\leq S_{20}(1.981001468867)$, followed by $U_6$ at
$C\cdot1.981001468867$.

The independent checker obtains logarithm intervals differently. After
writing $x=2^kz$ with $1\leq z\leq2$, it uses
\[
 \log z=2\sum_{j=0}^{N-1}\frac{y^{2j+1}}{2j+1}+R_N,
 \quad y=\frac{z-1}{z+1},\quad
 0\leq R_N\leq\frac{2y^{2N+1}}{(2N+1)(1-y^2)}.
\]
Here $N=40$ and $0\leq y\leq1/3$; the same formula encloses $\log2$.
For an exponential argument $0\leq x\leq1$, the checker uses $S_{16}(x)$
and the upper tail
$x^{16}/(16!(1-x/17))$. These bounds also certify the signs reported in
Section~\ref{sec:verification} without floating-point acceptance tests.

\clearpage
\input{data/certificate-bounds}

\end{document}

%% file: data/inventories.tex
\begin{tabular}{rr@{\qquad\qquad}rr}
\toprule
\multicolumn{2}{c}{Charged blocks $\mathcal P$} & \multicolumn{2}{c}{Target jobs $\mathcal T$} \\
Rank & Count & Rank & Count \\
\midrule
1 & 41946 & 4 & 504 \\
2 & 6984 & 7 & 126 \\
3 & 8892 & 8 & 846 \\
4 & 2646 & 9 & 5868 \\
5 & 648 & 17 & 448 \\
8 & 144 & 25 & 448 \\
16 & 448 & 33 & 448 \\
24 & 448 & 41 & 448 \\
32 & 448 & 49 & 448 \\
40 & 448 & 57 & 448 \\
48 & 448 & 65 & 448 \\
56 & 448 & 73 & 224 \\
64 & 448 & 145 & 224 \\
72 & 224 & --- & --- \\
\bottomrule
\end{tabular}

%% file: data/certificate-bounds.tex
\begingroup
\small
\setlength{\tabcolsep}{9pt}
\begin{longtable}{rrrr}
\caption{Exact rational enclosures for the normalized moment terms. The charged-block rows give upper bounds; the target rows give lower bounds. $N$ is the number of terms in the logarithm test described in the text.}\label{tab:bounds}\\
\toprule
Rank $r$ & Logarithm bound $\ell_r$ & $N$ & Moment bound $b_r$ \\
\midrule
\endfirsthead
\toprule
Rank $r$ & Logarithm bound $\ell_r$ & $N$ & Moment bound $b_r$ \\
\midrule
\endhead
\bottomrule
\endfoot
\multicolumn{4}{l}{\emph{Charged blocks: upper bounds}} \\
\addlinespace[2pt]
1 & 4.976733742421 & 29 & 293.663682394904 \\
2 & 4.283586561861 & 27 & 97.585379104181 \\
3 & 3.878121453753 & 26 & 186.139872110444 \\
4 & 3.590439381301 & 25 & 73.788927475228 \\
5 & 3.367295829987 & 24 & 22.573229270017 \\
8 & 2.897292200741 & 23 & 8.014652284989 \\
16 & 2.204145020181 & 21 & 49.764660119895 \\
24 & 1.798679912073 & 19 & 74.555635518839 \\
32 & 1.510997839621 & 18 & 99.321181555482 \\
40 & 1.287854288307 & 17 & 124.067835657887 \\
48 & 1.105532731513 & 16 & 148.799445098402 \\
56 & 0.951382051686 & 15 & 173.518550400695 \\
64 & 0.817850659061 & 15 & 198.226956274002 \\
72 & 0.700067623405 & 14 & 111.463005749806 \\
\midrule
\multicolumn{4}{l}{\emph{Target jobs: lower bounds}} \\
\addlinespace[2pt]
4 & 3.590439381300 & 12 & 14.055033804805 \\
7 & 3.030823593365 & 11 & 6.138693311557 \\
8 & 2.897292200740 & 11 & 47.086082174304 \\
9 & 2.779509165084 & 11 & 367.291038096216 \\
17 & 2.143520398364 & 10 & 52.865271049404 \\
25 & 1.757857917552 & 10 & 77.652546021820 \\
33 & 1.480226180954 & 10 & 102.415450004130 \\
41 & 1.263161675716 & 9 & 127.160048134959 \\
49 & 1.084913444309 & 9 & 151.889974497414 \\
57 & 0.933682474586 & 9 & 176.607655012455 \\
65 & 0.802346472524 & 8 & 201.314825634830 \\
73 & 0.686274301272 & 8 & 113.006395319643 \\
145 & 0 & --- & 224 \\
\end{longtable}
\endgroup